\documentclass[runningheads]{llncs}
\usepackage{graphicx}
\usepackage{tikz}
\usepackage{tkz-graph}
\usepackage{bbm}
\usepackage{booktabs}
\usepackage{listings}
\usepackage{amsfonts}
\usepackage{amsmath}
\usepackage{wrapfig}
\usepackage{pgfplots}
\usepackage{pgfplotstable}
\usepackage{adjustbox}
\usepackage{hyperref}

\newtheorem{mydef}{Definition}

\begin{document}
\title{Increasing Scalability of Process Mining using Event Dataframes: How Data Structure Matters}
\titlerunning{Increasing Scalability of Process Mining using Event Dataframes}
%
%
\author{Alessandro Berti\inst{1}\orcidID{0000-0003-1830-4013}}
%
%
\institute{Process and Data Science department, Lehrstuhl fur Informatik 9 52074 Aachen, RWTH Aachen University, Germany}
\maketitle              
\begin{abstract}
Process Mining is a branch of Data Science that aims to extract process-related information from event data contained in information systems, that is steadily increasing in amount.
Many algorithms, and a general-purpose open source framework (ProM 6), have been developed in the last years for process discovery, conformance checking, machine learning on event data.
However, in very few cases scalability has been a target, prioritizing the quality of the output over the execution speed and the optimization of resources. This is making progressively
more difficult to apply process mining with mainstream workstations on real-life event data with any open source process mining framework.
Hence, exploring more scalable storage techniques, in-memory data structures, more performant
algorithms is a strictly incumbent need. In this paper, we propose the usage of mainstream columnar storages and dataframes to increase the scalability of process mining.
These can replace the classic event log structures in most tasks, but require completely different implementations with regards to mainstream process mining algorithms.
Dataframes will be defined, some algorithms on such structures will be presented and their complexity will be calculated.
\end{abstract}

\section{Introduction}

Information systems support the execution of nowadays business processes, recording information about {\it events} related to the executions
of a business process. Relational (MySQL, Oracle, SQL Server, \ldots) and non-relational databases (MongoDB, ElasticSearch, \ldots) are used to store and query the event data.
Process Mining supports the extraction of process-related knowledge from information systems, targeting both systems supported by a workflow (some possible analyses are {\it bottleneck analysis}, {\it remaining time prediction}, {\it logical-temporal checking})
and systems where process executions are more loosely described (with {\it process discovery}, that is the automatic discovery of a process schema from the event data, and
{\it conformance checking} techniques). Several process mining algorithms have been developed for each goal.
Generally, a process mining analysis starts from the extraction of an {\it event log} from the information system. The event log is then imported into a process mining tool
(for example, the ProM6 Process Mining framework), and
several analyses to the same log can be applied. As an example, a popular process discovery technique is the inductive miner \cite{leemans2013discovering}, that extract a process model
(process tree or accepting Petri net) from the event log. Alignments are a popular conformance checking technique \cite{adriansyah2011conformance}, that enables compliance and performance analyses.

\begin{table*}[!t]
\label{tab:eventLogCharacterization}
\caption{Characterization of the event logs used in the assessment of the approaches. The loading time and the occupation of RAM have been measured against the XESLite log importer in the ProM6 process mining framework.
The measurement has been done on a Lenovo ThinkPad T470s notebook (I7-7550U, 16 GB RAM, DDR).}
\centering
\begin{tabular}{|l|cccc|cc|}
\hline
{\bf Log} & {\bf Events} & {\bf Cases} & {\bf Variants} & {\bf Classes} & {\bf Loading time(s)} & {\bf RAM(MB)} \\
\hline
roadtraffic & $561470$ & $15370$ & $231$ & $11$ & $9.60$ & $412$ \\
bpic2017\_o & $193849$ & $42995$ & $16$ & $8$ & $6.70$ & $421$ \\
bpic2017\_a & $1202267$ & $31509$ & $15930$ & $26$ & $49.9$ & $1203$ \\
bpic2018 & $2514266$ & $43809$ & $28457$ & $41$ & $39.4$ & $3219$ \\
bpic2019 & $1595923$ & $251734$ & $11973$ & $42$ & $29.8$ & $1435$ \\
\hline
\end{tabular}
\vspace{-7mm}
\end{table*}

A problem that surged in recent years is that the amount of information contained in modern information systems is steadily increasing. This pose challenges that were addressed only from a small
part of the process mining research. Taking into account a mainstream analytic workstation in 2019 (I7-8550U, 16GB DDR4, 3000 MB/s SSD sequential read), and a public available log as
the BPI Challenge 2018 log (2.5M events, 44K different process executions), importing the log through the efficient XESLite importer takes $50$ seconds and the resulting in-memory log structure occupies $3219$ MB of RAM.
More details about the features of the considered logs are contained in \autoref{tab:eventLogCharacterization}.
On the other hand, in recent years some efficient storage techniques have been proposed, that are called {\it columnar} because each column is saved separately inside the storage, including the popular
Apache Parquet and Apache ORC formats. Corresponding to the storage techniques,
an interesting class of data structures, the {\it dataframes}, has been implemented
to support fast filtering and aggregations of the underlying data.
With regards to open source process mining, at the best of our knowledge no previous usage of columnar storages in process mining has been described, while a previous scientific contribution \cite{cheng2019scalable} has used a dataframe structure
in order to increase the speed of computation in process mining, but without providing an abstraction of the data structure, or some considerations
on why dataframes are better suited to handle big amount of data.

In this paper, an abstraction of the notion of {\it dataframe} is provided (Pandas and Spark provide implementations of such structure),
explaining a basic implementation of dataframe, with the final goal to obtain a greater scalability in mainstream process mining operations.
Dataframes can directly ingest event data contained in columnar storages (like Apache Parquet and Apache ORC), and so they can exploit their advantages
(attribute selection at load time and efficient compression techniques).
Some complexity estimation for mainstream process mining operations, like filtering
and the Directly-Follows Graph computation, on the described dataframe implementation will be provided in order to show how dataframes
are suited for fast log exploration and filtering.
At the best of the authors knowledge, this is the first scientific paper combining in an holistic manner columnar storages and dataframes
to increase the scalability of process mining techniques.
Moreover, this is (at the best of the authors knowledge) the first abstraction of the dataframe concept, that has already very popular implementations in Pandas and Apache Spark.
The assessment (while avoiding, for fairness, direct comparison with the implementation of the classical event log structure,
done in different frameworks/languages) will show that popular dataframe implementations can handle big amount of data with a good execution time
and a reasonable memory occupation.

The paper is organized as follows.
In \autoref{sec:relatedWork}, the related work on storage techniques and in-memory/distributed processing and benchmarking is presented.
In \autoref{sec:background}, the concepts necessary to understand the paper will be introduced.
In \autoref{sec:complexity}, the complexity of the operations on classic event logs will be presented.
In \autoref{sec:eventDataframes}, the dataframe concept will be introduced along with some transformation functions, a conversion technique from an event log structure to a dataframe structure,
a simple implementation and a calculation of the complexity of operations on dataframes.
In \autoref{sec:assessment}, a dataframe implementation in the Python language (Python) will be assessed on real-life and simulated event logs.
Eventually, \autoref{sec:conclusion} concludes the paper.

\section{Related Work}
\label{sec:relatedWork}

{\it The XES standard}: 
several formats have been proposed during the years for the standard storage of event logs in process mining. The IEEE standard is XES \cite{verbeek2010xes}, for which different implementations exist
in the ProM6 process mining framework. Among noticeable implementations, we can cite XESLite \cite{mannhardt2016xeslite}, that provides a memory-efficient handling of event logs, while \cite{van2015relational}
supports the XES standard on relational databases, albeit with a performance deficit, and DB-XES \cite{syamsiyah2016db} that use relational databases to support some intermediate calculations.
OpyenXES \cite{valdiviesoopyenxes} took XES in an open source Python implementation, and the PM4Py Python process mining library \cite{berti2019process} followed obtaining a full certification. \\
{\it Benchmarking of process mining algorithms}:
the work \cite{augusto2018automated} contains an extensive benchmarking of some process mining algorithms. The evaluation highlight some problems of available techniques, including the lack of scalability and a strong divergence in their performance with respect to the different quality metrics.
In \cite{leemans2018scalable}, a variant of the popular inductive miner process discovery algorithm (based only on the directly-follows graph) is explained, along with extensive evaluation. \\
{\it Process Mining over Big Data}:
an approach of process mining over big data triples stores is contained in \cite{azzini2013consistent}.
A ProM 6 plug-in, enabling process mining on the Apache Hadoop big data platform, has been described in \cite{hernandez2015handling}.
In \cite{van2015processes}, some motivations on connecting process mining with the big data world are explained.
A distributed compliance monitoring of business processes over Mapreduce architectures is described in \cite{loreti2017distributed}.
An approach of process mining over the Apache Spark framework is described in \cite{cheng2019scalable}. \\
{\it Row- and Columnar- based approaches in Data Mining}:
in \cite{abadi2008column}, a comprehensive evaluation of column-stores versus row-stores is provided.
The implementation of a modern column-oriented database system is described in \cite{abadi2013design}.

\section{Background}
\label{sec:background}

\subsection{Event Logs: Definition and Considerations}

In this section, the notion of event log, i.e., a collection of events recorded in the execution of a business process that is the starting point of many process mining analysis, is formally introduced.
For the following definition, let $\mathcal{U}_E$ be the universe of events,
$\mathcal{U}_C$ be the universe of case identifiers,
$\mathcal{U}_A$ be the universe of activities (names referring to a particular step of a process),
$\mathcal{U}_{\textrm{attr}}$ be the universe of attribute names (all the names of the attributes that can be related to an event),
$\mathcal{U}_{\textrm{val}}$ be the universe of attribute values (all the possible values for attributes).

\begin{mydef}[Classical Event Log]
\label{def:classicalEventLogDefinition}
A log is a tuple $L = (C_I, E, A, \textrm{case\_ev}, \allowbreak \textrm{act}, \allowbreak \textrm{attr}, \leq)$ where
$C_I \subseteq \mathcal{U}_C$ is a set of case identifiers,
$E \subseteq \mathcal{U}_E$ is a set of events,
$A \subseteq \mathcal{U}_A$ is the set of activities,
$\textrm{case\_ev} \in C_I \rightarrow \mathcal{P}(E) \setminus \{ \emptyset \}$ maps case identifiers onto set of events (belonging to the case),
$\textrm{act} \in E \rightarrow \mathcal{U}_A$ maps events onto activities,
$\textrm{attr} \in E \rightarrow (\mathcal{U}_{attr} \not\rightarrow \mathcal{U}_{val})$ maps events onto a partial function assigning values to some attributes,
and $\leq ~ \subseteq E \times E$ defines a total order on events.
\end{mydef}

This classical event log notion matches the XES storage format \cite{verbeek2010xes}, that is the common source of information
for process mining tools like Disco, ProcessGold, Celonis, QPR, Minit, \ldots
An example attribute of an event $e$ is the timestamp $\textrm{attr}(e)(time)$ which refer to the time the event happened.
While, in general, an event belongs to a single case, in Def. \ref{def:classicalEventLogDefinition}
the function case\_ev might be such that cases share events.

An important consideration is that for an event $e \in E$, applying $\textrm{attr}$, the resulting mapping $\mathcal{U}_{attr} \not\rightarrow \mathcal{U}_{val}$ is a partial function.
This means that not all the attributes are valued for all the events, e.g., there might be some attributes that are specific for a single event.

\subsection{Directly-Follows Graph}
\label{sec:dfgDefinition}

The Directly-Follows Graph (DFG) is a basic representation of the activities of the process, that are connected with a directed if they follow each other
in at least an instance of the event log. The DFG is usually annotated by the number of occurrences for each edge (corresponding
to the number of occurrences of the log where an activity has been followed by another activity in the log).

The concept of a Directly-Follows Graph is formally introduced in the following definition:

\begin{mydef}[DFG]
A Directly-Follows Graph is a weighted directed graph:
$\textrm{DFG} = (N_{\textrm{DFG}}, E_{\textrm{DFG}}, c_{\textrm{DFG}})$
where $N_{\textrm{DFG}}$ (the nodes) are the activities, and $E_{\textrm{DFG}} \subseteq N_{\textrm{DFG}} \times N_{\textrm{DFG}}$ is the set of all the edges between activities that happened in direct succession,
and $c_{\textrm{DFG}} : E_{\textrm{DFG}} \rightarrow \mathbb{R}^{+}$ is the count function that aims to represent how many times two different activities happened in direct succession.
\end{mydef}

Let $\textrm{start}_{\textrm{DFG}} \subseteq N_{\textrm{DFG}}$ be the subset of start activities in the log,
and $\textrm{end}_{\textrm{DFG}} \subseteq N_{\textrm{DFG}}$ be the subset of end activities in the log. Then, the DFG structure
can be converted into a Petri net useful for conformance checking purposes using the method described in \cite{leemans2019directly}.
Moreover, a variant of the Inductive Miner (IMDF) is described in \cite{leemans2018scalable}.

\subsection{Row- and Columnar-based Storages}

\begin{figure*}[!t]
\centering
\includegraphics[width=345px]{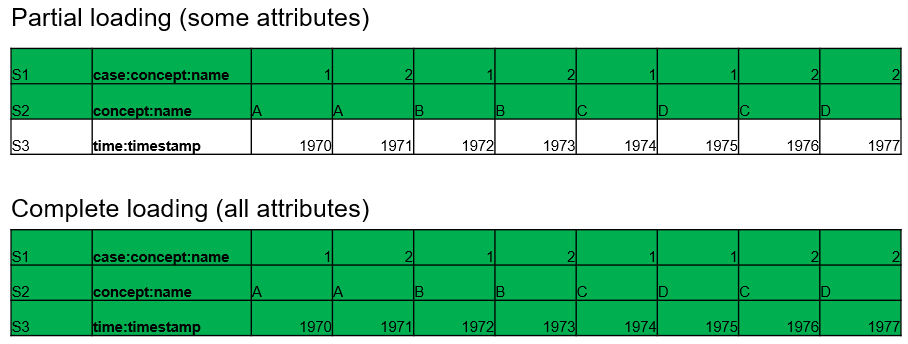}
\caption{Example selective loading of attributes. For most process discovery and conformance checking algorithms,
loading only the case ID ({\it case:concept:name}) and the activity ({\it concept:name}) is enough. So, why loading all the attributes if they are not needed?
For log exploration purposes, instead, it may be useful to load all the attributes (in order to filter interesting behavior).}
\label{fig:attributeSelection1}
\vspace{-5mm}
\end{figure*}

This section will introduce row- and columnar-based storages. While the remainder of the paper will be mainly focused on columnar storages and data structures, it is interesting to see how modern
row-based storages cope against the columnar storages when saving process mining event logs.

Row-based storage systems are optimized to read logs ``by rows'', ``by events'': while parsing, the row is considered in its entirety, and all the values for all the attributes of the parsed event
are read at first glance. A well-known row-based storage format is the Avro format\footnote{\url{https://avro.apache.org/}}. Each Avro file contains an header, where the parsing schema is defined,
and one or more data blocks (containing one or more rows). For process mining purposes, the event log can be transformed into a list of events, and then each event serialized into a row.

Column-based storage systems are optimized to read logs by ``by columns'', making possible to choose which attributes are needed before starting to load the file (see \autoref{fig:attributeSelection1}). In this way, only the values of some attributes
are parsed, making the load operation faster. Since each column of the file contains data of the same format (integer, string, ...), more effective compression techniques can be deployed.
Among the compression algorithms supported by Parquet, there is the {\it Snappy} compression\footnote{\url{https://en.wikipedia.org/wiki/Snappy_(compression)}} and the {\it Gzip} compression\footnote{\url{https://en.wikipedia.org/wiki/Gzip}}. Snappy is more fast in compressing/decompressing, while Gzip obtains better compression ratios at the expense of performance.
Apache Parquet\footnote{\url{https://parquet.apache.org/}} is supported by a large number of big data frameworks (Apache Hive, Apache Drill, Apache Impala, Apache Pig, Apache Spark, Cascading, \ldots).
Apache ORC\footnote{\url{https://orc.apache.org/}} is an alternative in the columnars. In \autoref{tab:sizeLogDisk1}, the size on disk of the logs stored in the different formats is reported.
The Apache Parquet format with the Gzip compression achieves the best results in terms of log size on disk, while Parquet with the Snappy compression remains very efficient while avoiding the compression/decompression performance deficit of Gzip.

\begin{table*}[!t]
\label{tab:sizeLogDisk1}
\caption{Size on disk of logs stored in XES, CSV, Avro, Parquet (with Snappy compression), ORC, XES (applying Gzip compression to the file), Parquet (applying Gzip compression to the columns).
The Gzip compression leads to a performance hit since the file and the columns need to be uncompressed before loading. It should be noted that, while in the CSV/ORC/Parquet formats only the event data is contained in the file, XES contains some additional metadata and information.}
\centering
\begin{tabular}{|l|ccccc|cc|}
\hline
{\bf Log} & {\bf S.XES} & {\bf S.CSV} & {\bf S.Avro} & {\bf S.Parq.(S)} & {\bf S.ORC} & {\bf S.XES(G)} & {\bf S.Parq.(G)} \\
\hline
roadtraffic & $181$MB & $47$MB & $34$MB & $6$MB & $3.9$MB & $3.3$MB & $3.6$MB \\
bpic2017\_o & $107$MB & $36$MB & $34$MB & $7$MB & $5.5$MB & $5.5$MB & $4.7$MB \\
bpic2017\_a & $565$MB & $205$MB & $184$MB & $29$MB & $22.0$MB & $29.1$MB & $17.1$MB \\
bpic2018 & $1901$MB & $1505$MB & $1955$MB & $167$MB & $146$MB & $155$MB & $96$MB \\
bpic2019 & $711$MB & $434$MB & $462$MB & $14$MB & $10.1$MB & $16.5$MB & $8.6$MB \\
\hline
\end{tabular}
\vspace{-5mm}
\end{table*}

\section{Approach: Complexity of operations on Attribute Maps}
\label{sec:complexity}

The aim of this section is to present the complexity of retrieval and of the Directly-Follows Graph calculation on top of classical event logs.
This will permit to compare the complexity of such operations against an implementation of the dataframe, that is introduced in \autoref{sec:eventDataframes}.

In \autoref{sec:relatedWork}, the current standard-de-facto implementations of event logs, based on the XES standard, have been presented.
The $\textrm{attr}$ function is usually implemented to return $\mathcal{U}_{attr} \not\rightarrow \mathcal{U}_{val}$ as an associative array data structure
(for example, the plain XES uses the \emph{HashMap} and XESLite uses the \emph{THashMap} (Trove map implementation)).
Hash maps are efficient in insertion and retrieval, with an average complexity of $O(1)$, but a worst case of $O(M)$ (where $M$ is the number of attributes of the specific event).
This means that the complexity of the retrieval of the entire set of values of a single event attribute, on all the events of the log, is in the worst case $O(N*M)$
(where $N$ is the number of events in the event log and $M$ is the maximum number of different attributes that any event has). The complexity in the average case is $O(N)$.
This holds also for common problems in process mining like filtering the log on specific attribute values, or extracting the Directly-Follows Graph (see \autoref{sec:dfgDefinition}).

The calculation of the Directly-Follows Graph could be performed iterating over the cases of the log. This has an average complexity of $O(N)$
since a single pass is enough and the storage of the edges/counts in a map on the log has an average linear complexity, but in the worst case
(where all the events belongs to a single case and the worst case of a map is encountered) has complexity of $O(N^2)$.

\begin{figure*}[!t]
\centering
\includegraphics[width=245px]{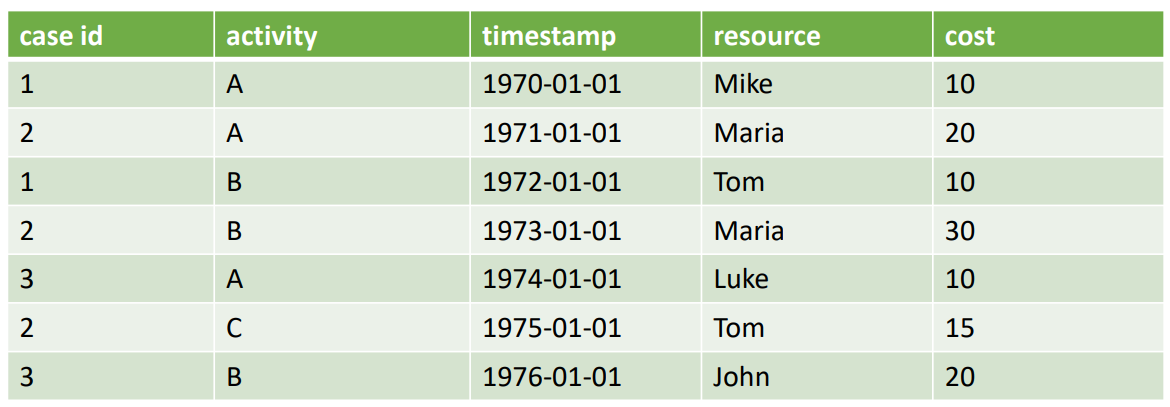}
\caption{Representation of a dataframe structure. In this table, the columns are corresponding to the names of the attributes.
}
\label{fig:exampleDataframe}
\vspace{-5mm}
\end{figure*}

\section{Approach: Dataframes}
\label{sec:eventDataframes}

This section introduces the techniques that are necessary to do process mining on top of a different structure, i.e. the {\it dataframe}.
Dataframes can ingest without hassle data contained in popular columnar formats like Apache Parquet or Apache ORC, that were introduced in
\autoref{sec:background}.

The section is organized as follows: in \autoref{sec:eventDfDefinition}, the abstraction of the dataframe structure is presented.
In \autoref{sec:conversion}, a basic implementation of a dataframe is presented, along with some conversion techniques from an event
log structure to a dataframe abstraction. In \autoref{sec:transformation}, some basic operations (that are useful for process mining purposes)
on a dataframe structure are introduced. In \autoref{sec:dfgOnDataframes}, some algorithms to calculate the Directly-Follows Graph
on dataframes are presented.

\subsection{Definition of Dataframes}
\label{sec:eventDfDefinition}

To introduce the concept of dataframe, one could think to a set of attribute names (columns) associated to the values (of the attribute) for each
row (that is an event). An example representation is contained in Fig. \ref{fig:exampleDataframe}.

\begin{mydef}[Dataframes]
Let $I \subset \mathbb{N}$ be a set of indexes, $N \subset \mathcal{U}_{\textrm{attr}}$ be a set of attribute names, $T \subset \mathcal{U}_{\textrm{types}}$ be a set of attribute types, $V \subset \mathcal{U}_{\textrm{val}}$ be a set
of attribute values for which a partial order $\leq_V$ exists. A {\it dataframe} is a tuple $D = (I, N, T, V, \chi^{\textrm{val}}_i, \chi^{\textrm{type}})$ where:
\begin{itemize}
\item $\chi^{\textrm{val}}_i : I \times N \rightarrow V \cup \{ \epsilon \}$ assigns to an index $i \in I$ and an attribute name $n \in N$ a value in $V \cup \{ \epsilon \}$.
\item $\chi^{\textrm{type}} : N \rightarrow T$ assigns to an attribute name $n \in N$ a type in $T$.
\end{itemize}
An assumption that is being made is that attributes $n_{\textrm{case}}$ (the case) and $n_{\textrm{act}}$ (the activity) exist, such that
for all $i \in I$, $\chi^{\textrm{val}}_i(n_{\textrm{case}}) \neq \epsilon$ and $\chi^{\textrm{val}}_i(n_{\textrm{act}}) \neq \epsilon$.
\label{def:dataframes}
\end{mydef}

In the following, we consider $\mathcal{U}_D$ to be the {\it universe} of dataframes,
and $\chi^{\textrm{val}} : N \rightarrow I \times (P \cup \{ \epsilon \})$ is a function that associates to an attribute name the set of all indexed values, i.e., for $n \in N$:
$\chi^{\textrm{val}}(n) = \cup_{i \in I} \{ (i, \chi^{\textrm{val}}_i(n)) \}$.
Hence, if we take the second element of the couples, we get the set of all attribute values for $n$, that we indicate with $\pi_2(\chi^{\textrm{val}}(n))$.
Moreover, given a dataframe $D \in \mathcal{U}_D$, the function $\pi_I : \mathcal{U}_D \rightarrow \mathcal{P}(\mathbb{N})$, $\pi_I((I, N, T, V, \chi^{\textrm{val}}_i, \chi^{\textrm{type}})) = I$
is defined.

\subsection{Simple Implementation of a Dataframe, Conversion from Event Logs to Dataframes}
\label{sec:conversion}

The data structure defined in \autoref{sec:eventDfDefinition} is actually an abstraction of the one that is implemented in Pandas (Python) and
Apache Spark. In this section, we want to describe a simple implementation of the dataframe concept, assuming that the name of the attributes
is represented as a string. The proposed implementation is the following: {\bf Map$<$String, ArrayList$<$Object$>>$} where:
{\bf Map} is any associative array implementation\footnote{\url{https://en.wikipedia.org/wiki/Associative_array}} and
{\bf ArrayList} is any dynamic array implemented\footnote{\url{https://en.wikipedia.org/wiki/Dynamic_array}}.
In the previous data structure, all the values for a particular attribute can be retrieved from the map as an array list of objects. In this data structure,
the value in position $i$ is the value of the attribute for the event at position $i$ in the dataframe.

Converting from an event log
$L = (C_I, E, A, \textrm{case\_ev}, \allowbreak \textrm{act}, \allowbreak \textrm{attr}, \leq)$
to a dataframe
$D = (I, N, T, V, \chi^{\textrm{val}}_i, \chi^{\textrm{type}})$
requires the assumption that an attribute $n_{\textrm{case}} \in \mathcal{U}_{\textrm{attr}}$ exists
in the domain of $\textrm{attr}(e)$ for each event $e \in E$. Since $\leq$ (see \autoref{def:classicalEventLogDefinition}) is a total order on $E$,
$E$ can be viewed as a sequence of events,
$E = \{ e_1, e_2, \ldots, e_n \}$,
such that $e_i \leq e_j$ if and only if $i \leq j$. 
The conversion is such that $N = \mathcal{U}_{attr}$ is the set of attributes of the dataframe,
and $T = \{ o \}$ can be defined to contain a single (object) type. Then,
$V = \cup_{e \in E} \textrm{Im} ~ \textrm{attr}(e)$, $\chi^{\textrm{type}}(n) = t ~ \forall n \in N$, and
$$\chi^{\textrm{val}}_i(n) =
\begin{cases}
\textrm{attr}(e_i)(n) & {\textrm if} ~ n \in \textrm{Dom} ~ \textrm{attr}(e) \\
\epsilon & {\textrm if} ~ n \not\in \textrm{Dom} ~ \textrm{attr}(e)
\end{cases}
$$

\subsection{Transformation functions}
\label{sec:transformation}

We introduce the following functions in order to be able to perform process mining operations on top of a dataframe
(as the calculation of Directly-Follows Graphs).
These functions transform a/some dataframe(s) into another dataframe.
\begin{itemize}
\item {\it Projection on a given expression}: given a dataframe $D = (I, N, T, V, \chi^{\textrm{val}}_i, \chi^{\textrm{type}})$, a set of attribute names $S \subseteq N$ and a function $f: I \times \mathcal{P}(N) \rightarrow \{0 ,1 \}$ (selective function based on the attribute values),
a projection function is defined as
$\textrm{proj}((I, N, T, V, \chi^{\textrm{val}}_i, \chi^{\textrm{type}}), S, f) = (I', N, T, V, \chi'^{\textrm{val}}_i, \chi^{\textrm{type}})$
where $I' = \{ i \in I, f(i, S) = 1 \}$ and $\chi'^{\textrm{val}}_i(n) = \chi^{\textrm{val}}_i(n)$ ~ for all $i \in I'$ and $n \in N$.
\item {\it Grouping function}: the grouping function is defined given a dataframe $D = (I, N, T, V, \chi^{\textrm{val}}_i, \chi^{\textrm{type}}) \in \mathcal{U}_D$
and an attribute $n_0 \in N$ as
$$\textrm{group}(D, n_0) = \bigcup_{v \in \pi_2(\chi^{\textrm{val}}(n_0))} \textrm{proj}(D, n_0, \{ v \})$$
by construction, $\textrm{proj}(D, n_0, \{ v \}) \cap \textrm{proj}(D, n_0, \{ v' \}) = \emptyset$ if $v \neq v'$,
so the sets are pairwise disjoint.
\item {\it Shift function}: $\textrm{shift}((I, N, T, V, \chi^{\textrm{val}}_i, \chi^{\textrm{type}})) = (I', N, T, V, \chi'^{\textrm{val}}_i, \chi^{\textrm{type}})$ where
$I' = \{ i - 1 \arrowvert i \in I \}$, $\chi'^{\textrm{val}}_{i - 1}(n) = \chi^{\textrm{val}}_{i}(n)$ ~ $\forall i \in I, n \in \mathbb{N}$.
\item {\it Concatenation function}: for $D_1 = (I_1, N_1, T_1, V_1, \chi'^{\textrm{val}}_i, \chi'^{\textrm{type}}) \in \mathcal{U}_D$, $D_2 = (I_2, N_2, T_2, V_2, \chi''^{\textrm{val}}_i, \chi''^{\textrm{type}}) \in \mathcal{U}_D$, and $s \in \Sigma$ such that
$\{ n_2 \oplus s \arrowvert n_2 \in N_2 \} \cap N_1 = \emptyset$, the concatenation is defined as
$\textrm{concat}(D_1, D_2, s) = (I, N, T, V, \chi^{\textrm{val}}_i, \chi^{\textrm{type}})$
where $I = I_1 \cup I_2$, $N = N_1 \cup \{ n_2 \oplus s \arrowvert n_2 \in N_2 \}$, $T = T_1 \cup T_2$, $V = V_1 \cup V_2$,
$$
{\scriptscriptstyle 
\chi^{\textrm{val}}_i(n) = \begin{cases}
\chi'^{\textrm{val}}_i(n) & n \in N_1 \\
\chi''^{\textrm{val}}_i(n') & n = n' \oplus s, n' \in N_2 \\
\end{cases}
\quad
\chi^{\textrm{type}}(n) = \begin{cases}
\chi'^{\textrm{type}}(n) & n \in N_1 \\
\chi''^{\textrm{type}}(n') & n = n' \oplus s, n' \in N_2 \\
\end{cases}
}
$$
\item {\it Sorting function}: let $D = (I, N, T, V, \chi^{\textrm{val}}_i, \chi^{\textrm{type}})$ be a dataframe and $n \in N$ an attribute name. Let $b_n : I \rightarrow \{ 1, \ldots, \arrowvert I \arrowvert \}$ a bijection such that:
\begin{itemize}
\item If $\chi^{\textrm{val}}_{i}(n) \leq_V \chi^{\textrm{val}}_{j}(n)$ according to the partial order defined on $V$, then $b_n(i) \leq_{\mathbb{N}} b(j)$.
\item If $\chi^{\textrm{val}}_{i}(n) \leq_V \chi^{\textrm{val}}_{j}(n) \wedge \chi^{\textrm{val}}_{j}(n) \leq_V \chi^{\textrm{val}}_{i}(n)$, then $b_n(i) \leq_{\mathbb{N}} b(j) \iff i \leq_{\mathbb{N}} j$.
\end{itemize}
With this definition, $\textrm{sort}(D, n, b) = (\{ 1, \ldots, \arrowvert I \arrowvert \}, N, T, V, \chi'^{\textrm{val}}_i, \chi^{\textrm{type}})$ could be introduced as
$\chi'^{\textrm{val}}_{b_n(i)}(n) = \chi^{\textrm{val}}_{i}(n)$.
\item {\it String attribute merging}: let $D = (I, N, T, V, \chi^{\textrm{val}}_i, \chi^{\textrm{type}})$ be a dataframe, then a new attribute $n \in \mathcal{U}_{\textrm{attr}}$ of type $\textrm{str} \in \mathcal{U}_T$ could be inserted into the dataframe as merge of the values of two (existing) attributes $n_1, n_2 \in N$
(separated by a string delimiter $s$)
to obtain a new dataframe
$$\textrm{mergstrv}(D, n, n_1, n_2, s) = (I, N', T, V', \chi'^{\textrm{val}}_i, \chi'^{\textrm{type}})$$
where: $N' = N \cup \{ n \}$ and $V' = V \bigcup \cup_{i \in I} \{ \chi^{\textrm{val}}_i(n_1) \oplus s \oplus \chi^{\textrm{val}}_i(n_2) \}$
\end{itemize}

\subsection{Possibility to calculate the Directly-Follows Graph on top of dataframes}
\label{sec:dfgOnDataframes}

\begin{figure}
\vspace{-4mm}
\centering
\includegraphics[width=220px]{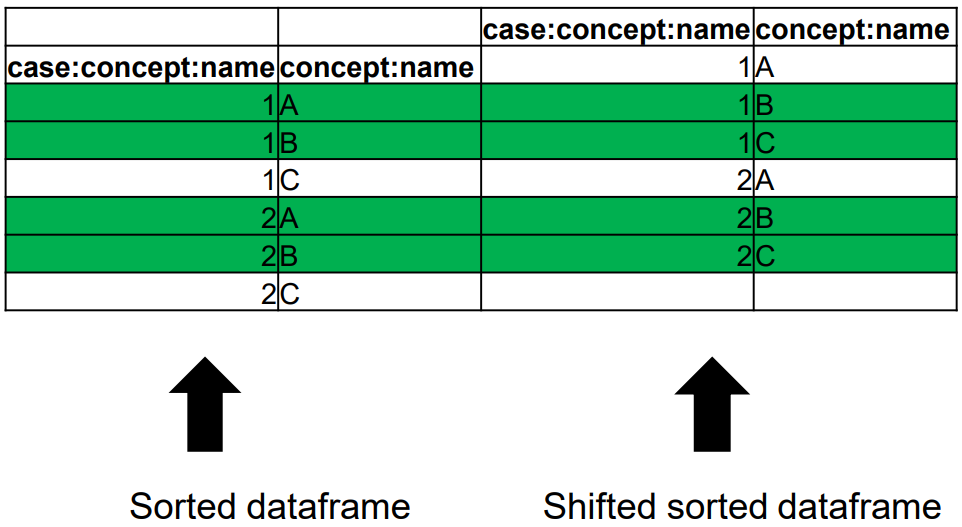}
\vspace{-1mm}
\caption{Example of the calculation of the Directly-Follows Graph on top of dataframes, using the shifting and counting method.}
\label{fig:reprShiftCounting}
\vspace{-5mm}
\end{figure}

In this section, we want to show that it's actually possible to calculate the Directly-Follows Graph (DFG; see \autoref{sec:dfgDefinition}) on top of the dataframe structures.
Let $D = (I, N, T, V, \chi^{\textrm{val}}_i, \chi^{\textrm{type}})$ be a dataframe.
Let $n_{\textrm{case}} \in N$ be the case identifier attribute, and $n_{\textrm{act}} \in N$ be the attribute that identifies the activity.
The nodes of the DFG are simply the different values for the activities:
$\pi_2(\chi^{\textrm{val}}(n_{\textrm{act}}))$.
For the retrieval of the edges and the count, two different strategies are actually possible:
\begin{itemize}
\item {\it Map-reduce}: the dataframe is grouped using the $n_{\textrm{case}}$ attribute, and then the edges and the counts are retrieved as:
$$
{\scriptstyle 
E = \bigcup_{G \in \textrm{group}(D, n_{\textrm{case}})}
\bigcup_{(i, j) \in \pi_I(G)^2, i < j \wedge \not\exists k \in \pi_I(G), i < k < j
} \left \{ \left ( \chi^{\textrm{val}}_i(n_{\textrm{act}}), \chi^{\textrm{val}}_j(n_{\textrm{act}}) \right ) \right \}
}
$$
For $e \in E$,
$$
{\scriptstyle 
c(e) = \sum_{G \in \textrm{group}(D, n_{\textrm{case}})} \sum_{(i, j) \in \pi_I(G)^2, i < j \wedge \not\exists k \in \pi_I(G), i < k < j}
\mathbbm{1}_{(\chi^{\textrm{val}}_i(n_{\textrm{act}}), \chi^{\textrm{val}}_j(n_{\textrm{act}})) = e}
}
$$
In a map-reduce implementation, the grouping is performed and then different groups can be assigned to different workers (\emph{map} phase).
The single results are then merged in the \emph{reduce} phase to obtain the final result.
\item {\it Shifting and counting}: the strategy assumes that the dataframe is sorted 
the strategy could be explained as follows:
\begin{enumerate}
\item A shifted version of the dataframe is obtained using the $\textrm{shift}$ function.
\item Then, a concatenated dataframe is obtained as:
$D' = (I', N', T', V', \chi'^{\textrm{val}}_i, \chi'^{\textrm{type}}) = \textrm{concat}(D, \textrm{shift}(D), \textrm{"\_2"})$.
\item Then, if $n_{\textrm{case}}$, $n'_{\textrm{case}}$ are the attributes identifying the case respectively in $D$ and $D'$, a dataframe $D''$ could be obtained taking $S = \{ n_{\textrm{case}}, n'_{\textrm{case}} \}$,
and $f : I \times \mathcal{P}(N) \rightarrow \{ 0, 1 \}$ such that
$f(i, S) = \mathbbm{1}_{\chi'^{\textrm{val}}_i(n_{\textrm{case}}) = \chi'^{\textrm{val}}_i(n'_{\textrm{case}})}$,
as
$D'' = (I'', N', T', V', \chi''^{\textrm{val}}_i, \chi'^{\textrm{type}}) = \textrm{proj}(D', S, f)$.
In such way, only rows in the concatenated dataframe that have the same case ID are kept.
\item Eventually, a new dataframe with an additional column $n''_{\textrm{act}}$ can be defined starting from $D''$, if $n_{\textrm{act}}$, $n'_{\textrm{act}}$ are the attributes identifying the activity respectively in $D$ and $D'$, as:
$$D''' = (I''', N'', T'', V'', \chi'''^{\textrm{val}}_i, \chi''^{\textrm{type}}) = \textrm{mergstrv}(D'', n''_{\textrm{act}}, n_{\textrm{act}}, n'_{\textrm{act}}, \textrm{","})$$
\item The edges and the count of the DFG are:
$E = \pi_2(\chi'''^{\textrm{val}}(n''_{\textrm{act}}))$
and
$c: E \rightarrow \mathbb{N}, c(e) = \arrowvert \{ i \in I \arrowvert \chi'''^{\textrm{val}}_i(n''_{\textrm{act}}) = e \} \arrowvert$
\end{enumerate}
\end{itemize}
A representation of the ideas underlying the algorithm is contained in \autoref{fig:reprShiftCounting}.

\subsection{Complexity Estimation on Logs and Dataframe Implementations}

{\it Filtering}: Filtering on the attribute values of events is an ubiquitous operation in process mining, since only a part of the events of the log may be of interest for an analysis.
Even in the worst case, accessing the value of an attribute for an event has $O(1)$ complexity, hence filtering a dataframe with $N$ events has a complexity that is in the worst case equal to
$O(N)$. A comparison between the complexity of filtering on event logs and dataframe implementations is contained in \autoref{tab:complexityFiltering}.

\begin{table}
\vspace{-3mm}
\label{tab:complexityFiltering}
\caption{Comparison between the theoretical complexity of filtering on the attribute values of events on implementation of logs and dataframe structures (average and worst case).
Here, $N$ is the number of events in the log, and $M$ is the number of different names for the attributes.
Since the access to attribute values on classic event logs require access to a map, the complexity of filtering in the worst case is greater
than the complexity of filtering operations on dataframes.}
\begin{tabular}{|p{4.5cm}|cc|}
\hline
{\bf Implementation} & {\bf Complexity - Average} & {\bf Complexity - Worst} \\
\hline
Iteration on a classic event log & $O(N)$ & $O(N*M)$ \\
Dataframe & $O(N)$ & $O(N)$ \\
\hline
\end{tabular}
\vspace{-4mm}
\end{table}

{\it Directly-Follows Graph computation}: the computation of the DFG on top of classical event logs requires a single iteration on the log and the storage of the DFG edges in a map.
Hence, the complexity is $O(N)$ in the average case and $O(N^2)$ in the worst case where all the events belong to the same case and a lot of collisions happen in the map.
As seen in \autoref{sec:dfgOnDataframes}, there are two ways to calculate the DFG on top of dataframes: map-reduce, and ``shifting and counting''.

For the map-reduce approach, the complexity is still $O(N)$ in the average case because all the operations involved (grouping, mapping and reducing to the map of edges and counts) have a linear average complexity.
The reduce approach, in the worst case, can have $O(N^2)$ complexity when a lot of collisions inside the map happen.
However, the real-life map-reduce implementations on dataframes (Pandas, Spark) are disappointing, being slow in the map operation and/or hampered
by network traffic issues (when the computations are split between different nodes)

For the ``shifting and counting'' approach, the initial sort on the dataframe has a complexity (if algorithms like mergesort and heapsort are used)
equal to $O(N * log(N))$ in the average and worst case. Then, every remaining operation has linear complexity, hence excluding the sorting operation,
the complexity of the remaining operations is still $O(N)$ in the average case, and $O(N^2)$ in the worst (that happens when a lot of collisions
happen in writing the edges and their count in the map).

To recap, the main difference between classical event logs and dataframes is the way attributes at event level are accessed:
\begin{itemize}
\item On classical event logs, the attribute map of the event is retrieved, and a specific key of the map is accessed to retrieve the value of the attribute.
\item On dataframes, the list of values for the specific attribute is retrieved, and then a specific index (corresponding to the index of the row/event) is retrieved.
\end{itemize}

A comparison between the filtering complexity on classical event logs and dataframe implementations is contained in \autoref{tab:complexityDFG}.
\begin{table}
\vspace{-4mm}
\label{tab:complexityDFG}
\caption{Comparison between the theoretical complexity of DFG calculation on implementations of logs and dataframe structures (average and worst case).
Here, $N$ is the number of events in the log. The iteration, the map-reduce and the shifting approach (although, this one assumes that a sorting of the dataframe on the case identifier is done) share the same
average and worst case complexity.}
\begin{tabular}{|p{4.5cm}|cc|}
\hline
{\bf Implementation} & {\bf Complexity - Average} & {\bf Complexity - Worst} \\
\hline
Iteration on a classic event log & $O(N)$ & $O(N^2)$ \\
Dataframe - Map-Reduce & $O(N)$ & $O(N^2)$ \\
Dataframe - Shifting(sort by Case ID to be done) & $O(N*log(N))$ & $O(N^2)$ \\
Dataframe - Shifting (sort by Case ID already done) & $O(N)$ & $O(N^2)$ \\
\hline
\end{tabular}
\vspace{-4mm}
\end{table}

\section{Assessment}
\label{sec:assessment}

In this section, an assessment of the speed of a dataframe implementation (Pandas), that can ingest data from the Parquet columnar format
is done on some popular event logs (a list of logs and features is contained in \autoref{tab:eventLogCharacterization}).
While a direct comparison is avoided (it would be probably unfair to compare the ProM6 process mining framework,
the final goal is to show that implementations of the dataframe structure can handle well modern real-life event logs,
using the advantages provided by columnar log formats (higher compression ratios, possibility to choose the attributes at load time).
The results are obtained using the PM4Py python process mining library (\url{http://www.pm4py.org}), importing/exporting dataframes from/to Parquet files
(see \url{http://pm4py.pads.rwth-aachen.de/documentation/working-with-event-data/}) and using dataframe computation techniques offered by the library
(see \url{http://pm4py.pads.rwth-aachen.de/documentation/big-dataframe-management/}).
The methodology and the results of the assessment are contained in tables $5$ and $6$.
Some further evidence that a dataframe structure provides a significant support for process mining operations is contained in the PM4Py
web services\cite{berti2019webservices}. There, different handlers have been built for classic event logs (ingested from XES)
and dataframes (ingested from CSV or Parquet).

\begin{table}
\label{tab:results1}
\vspace{-4mm}
\footnotesize
\caption{Assessment of columnar storages and dataframes on real-life logs (available in the 4TU repository).
All the experiments (except the DFG calculation) are repeated 1) including all the attributes, 2) including only the case ID and the activity column.
The experiments are: the measurement of the size on disk (MB), the loading time (in seconds), the amount of RAM needed to store the Pandas dataframe,
the time (in seconds) needed to filter the dataframe keeping events of the most common activity, and the DFG calculation time (in seconds) using the shifting and counting algorithm.
The measurements have been done on a Lenovo ThinkPad T470s notebook (I7-7550U, 16 GB RAM, DDR) and show that columnar storages/dataframes can handle very well current real-life logs up
to 2.5 millions of events.
}
\centering
\begin{tabular}{|l|cc|cc|cc|cc|c|}
\hline
{\bf Log} & {\bf S.Parq.} & {\bf 2c} & {\bf LoadT.} & {\bf 2c} & {\bf RAM} & {\bf 2c}  & {\bf A.Filt.} & {\bf 2c} & {\bf DFG(2c)} \\
\hline
roadtraffic & $6$MB & $4$MB & $0.42$s & $0.18$s & $73$MB & $13$MB & $0.30$s & $0.12$s & $0.24$s \\
bpic2017\_o & $7$MB & $1.7$MB & $0.32$s & $0.13$s & $24$MB & $4$MB  & $0.11$s & $0.04$s & $0.09$s \\
bpic2017\_a & $29$MB & $5.7$MB & $1.93$s & $0.33$s & $184$MB & $28$MB & $0.60$s & $0.16$s & $0.59$s \\
bpic2018 & $167$MB & $11.8$MB & $4.88$s & $0.43$s & $786$MB & $58$MB & $2.68$s & $0.36$s & $1.17$s \\
bpic2019 & $14$MB & $3.1$MB & $2.92$s & $0.59$s & $234$MB & $24$MB & $1.37$s & $0.37$s & $0.88$s \\
\hline
\end{tabular}
\vspace{-7mm}
\end{table}

\begin{table}
\label{tab:results3}
\vspace{-3mm}
\footnotesize
\caption{Assessment of columnar storages and dataframes on synthetic big logs (available at the address \url{https://drive.google.com/open?id=1VgcELM6NqJxWxs3kC2jOZbYpK2F1w9tV}).
The experiments are: the measurement of the size on disk (MB), the loading time (in seconds), the amount of RAM needed to store the Pandas dataframe, the time (in seconds) needed to filter the dataframe keeping events of the most common activity, and the DFG calculation time (in seconds) using the shifting and counting algorithm.
The measurements have been done on a Lenovo ThinkPad T470s notebook (I7-7550U, 16 GB RAM, DDR) and show that dataframes can efficiently handle big logs of $10^6$ cases and $10^7$ events.
}
\centering
\begin{tabular}{|l|ccc|cccc|}
\hline
{\bf Log} & {\bf N.Cas.} & {\bf N.Ev.} & {\bf Disk(MB)} & {\bf Load} & {\bf RAM(MB)} & {\bf Filt.} & {\bf DFG}\\
\hline
L1 & $1000000$ & $7003896$ & $78$ & $2.60$s & $300$ & $0.73$s & $4.39$s \\
L2 & $2000000$ & $14001120$ & $157$ & $4.65$s & $650$ & $1.36$s & $8.85$s \\
L3 & $3000000$ & $20999048$ & $237$ & $7.33$s & $950$ & $2.00$s & $13.17$s \\
L4 & $4000000$ & $34989340$ & $396$ & $12.66$s & $1600$ & $3.38$s & $22.49$s \\
L5 & $5000000$ & $48988692$ & $554$ & $17.08$s & $2200$ & $4.99$s & $33.48$s \\
\hline
\end{tabular}
\vspace{-3mm}
\end{table}

\section{Conclusion}
\label{sec:conclusion}

In this paper, a combination of columnar storages (like Apache Parquet and Apache ORC) and dataframes (popular implementations are Pandas and Apache Spark)
is used in order to increase the scalability of process mining techniques.
All the described techniques are implemented in the PM4Py library\cite{berti2019process} and the PM4Py web services\cite{berti2019webservices}.
The advantages of exhisting columnar formats are the efficient compression techniques that are available when data of the same attribute (having always the same type)
is saved in the same record, and the attribute selection at load time (that helps to avoid loading unnecessary attributes).
For dataframes, while some popular implementations already exist, this paper (at the best of the authors knowledge) provides the first abstraction of the concept of dataframe,
describing a set of operations that are feasible on the data structure and how to obtain a conversion from classical event logs to a dataframe structure.
The provision of such structure, and of a basic implementation of dataframe, is important to understand the differences with the classical event log and to provide a set
of algorithms on top of dataframes. Theoretical complexity is evaluated for two mainstream process mining techniques, for both algorithms on classical event logs implementations and the basic dataframe implementation.

The theoretical complexity estimation and the assessment shows that dataframes, combined with columnar storages, can provide a way to efficiently
handle millions of events. This is thanks to specific algorithms that are defined on the dataframe structure (filtering at the event level, and DFG calculation).
Hence, the techniques described in this paper seem suited to increase process mining scalability. Direct comparison with the event log structure is avoided since
the underlying frameworks/languages are different (ProM6/Java and PM4Py/Python), hence fairness of the analysis would be at risk.
However, to make use of the dataframe structure, specific algorithms are needed. This require a complete re-engineering of the process mining techniques.
For example, filtering at case level requires custom techniques that could be less efficient than the ones available for classical event logs.
Authors have already taken into PM4Py several dataframe-specific techniques (filtering at event and case level, statistics for cases/variants, LTL Checker).
Another limitation is the fact that, while in the XES standard, some metadata and additional information related to the process is stored inside the XML file,
the possibility to store them in columnar storages like Apache Parquet and Apache ORC has not be cleared in the pages of this paper.

XES is, and will be, an IEEE standard format that ensures interoperability between different process mining tools. It should also be said that
Parquet is a format that in the latest years is very used in data extraction from information systems, hence having an efficient structure
on top of columnars improves the integration with databases and processing techniques.
Overall, this paper is an introduction to columnar storages and dataframes in the field of process mining. Some further development
should be expected on the data partitioning for sampling and the distributed application of process mining algorithms.

\bibliographystyle{splncs04}
\bibliography{bigdataparquet}

\end{document}